\documentclass{article}

\usepackage{arxiv}

\usepackage[utf8]{inputenc} 
\usepackage[T1]{fontenc}    
\usepackage{hyperref}       
\usepackage{url}            
\usepackage{booktabs}       
\usepackage{amsfonts}       
\usepackage{nicefrac}       
\usepackage{microtype}      
\usepackage{lipsum}
\usepackage{graphicx}
\usepackage[center]{caption}

\title{Fused Deep Convolutional Neural Network for Precision Diagnosis of COVID-19 Using Chest X-Ray Images}

\author{ Hussin K. Ragb, Ian T. Dover \\
  Department of Engineering\\
  School of Electrical and Computer Engineering\\
  Christian Brothers University\\
  Memphis, Tennessee\\
  \texttt{\{hragb, idover\}@cbu.edu} \\
   \And
 Redha Ali \\
  Department of Electrical and Computer Engineering\\
  University of Dayton\\
  300 College Park, Dayton, Ohio 45469 \\
  \texttt{almahdir1@udayton.edu} \\
}

\begin{document}
\maketitle

\begin{abstract}
With a Coronavirus disease (COVID-19) case count exceeding 10 million worldwide, there is an increased need for a diagnostic capability. The main variables in increasing diagnostic capability are reduced cost, turnaround or diagnosis time, and upfront equipment cost and accessibility. Two candidates for machine learning COVID-19 diagnosis are Computed Tomography (CT) scans and plain chest X-rays. While CT scans score higher in sensitivity, they have a higher cost, maintenance requirement, and turnaround time as compared to plain chest X-rays. The use of portable chest X-radiograph (CXR) is recommended by the American College of Radiology (ACR) since using CT places a massive burden on radiology services. Therefore, X-ray imagery paired with machine learning techniques is proposed a first-line triage tool for COVID-19 diagnostics. In this paper we propose a computer-aided diagnosis (CAD) to accurately classify chest X-ray scans of COVID-19 and normal subjects by fine-tuning several neural networks (ResNet18, ResNet50, DenseNet201) pre-trained on the ImageNet dataset. These neural networks are fused in a parallel architecture and the voting criteria are applied in the final classification decision between the candidate object classes where the output of each neural network is representing a single vote.  Several experiments are conducted on the weakly labeled COVID-19-CT-CXR dataset consisting of 263 COVID-19 CXR images extracted from PubMed Central Open Access subsets combined with 25 normal classification CXR images. These experiments show an optimistic result and a capability of the proposed model to outperforming many state-of-the-art algorithms on several measures. Using k-fold cross-validation and a bagging classifier ensemble, we achieve an accuracy of 99.7\% and a sensitivity of 100\%.
\end{abstract}


\section{Introduction}
In 2020, we have seen a daily rapid increase in the COVID-19 case count reported from 20,000
daily new cases in mid-March to nearly 140,000 new cases daily in late-June. It is
imperative that fast and efficient methods of testing be adopted to meet the increased demand.
LabCorp’s COVID-19 at-home test kit, a test swab that is taken from home then delivered to a 
laboratory setting, is currently listed at over \$100 (USD) and has a multi-day turnaround time
due to shipment and laboratory procedures. As a less expensive option, Quidel sells reverse
transcriptase-polymerase chain reaction (RT-PCR) assays at around \$20 (USD) per test to health
care providers for viral nucleic acid detection; however, while RT-PCR tests are considered the
gold standard for COVID-19 diagnosis, they suffer from low sensitivity and a turnaround time in
many cases exceeding 4 hours. 

RT-PCR testing has been adopted as the forerunner in the Coronavirus pandemic, but
these tests yield a sensitivity of 91\% in practice \cite{wong2020frequency}. CT images have been shown to have higher
sensitivity in some cases to detect COVID-19 infections in patients with false-negative RT-PCR
results with reported sensitivities as high as 97-98\% \cite{wong2020frequency}. Researchers from Wuhan, China
developed a deep learning model named COVID-19 Detection Neural Network (COVNet) using
chest CT images, dividing results into 3 class labels: COVID-19, community-acquired
pneumonia (CAP), and non-pneumonia. This network acquired a sensitivity of 90\% and
specificity of 96\% \cite{wang2020covid}. 

Nevertheless, while recent COVID-19 radiology literature has extensively explored the
use of CT imagery, there are several benefits to applying CXR imagery to COVID-19 testing:
CXR imagery is relatively inexpensive compared to alternatives, has a decreased risk of infection, and is less invasive. CXR equipment
is more widely available, especially in developing nations, and the maintenance requirement of
sanitization is less than that for CT equipment. While the baseline sensitivity for CXR is 69\%, \cite{jacobi2020portable,chung2020ct,ng2020imaging}
CXR can play a primary role of the initial screening for COVID-19. In many cases, for patients
with high CXR abnormalities, further CT testing may be redundant, thus reducing the burden on
radiology services worldwide. In this way, it is suggested that both CXR and CT be used as a
two-pronged approach to early COVID-19 detection whereas CXR can be used as a first-line
triage tool. 

From a pictorial perspective of radiological images, CXR is the less sensitive modality
with ground glass densities (GGO) easily detectable on CT but not CXR; CXR densities often
appear hazy while CT densities have a clear contrast. However, reticular opacities are often more
apparent on CXR than on CT. On baseline CXR, consolidations are the most common finding,
with COVID-19 and viral pneumonia producing lung opacities in multiple lobes as opposed to
one in bacterial pneumonia which tends to be unilateral. COVID-19 pneumonia has prominent
peripheral air space opacities which are readily identified by both CT and CXR with CT
reporting peripheral lung distribution in 33-86\% of cases \cite{chung2020ct,ng2020imaging}. Diffuse air space diseases such
as acute respiratory distress syndrome (ARDS) demonstrate similar patterns to CXR, but these
COVID-19 lung opacities rapidly evolve into a diffuse coalescent or consolidative pattern with
1-3 weeks of symptom onset \cite{zhou2020ct,bernheim2020chest}. One study reported that of 64 patients, consolidation was
found in (30/64, 47\%) of patients while GGO was found in (21/64, 33\%). Common distribution
locations included peripheral (26/64, 41\%) and lower zone (32/64, 50\%), with most showing
bilateral involvement (32/64, 50\%) \cite{wong2020frequency}. Although very uncommon, pleural effusions were found
in a small number of cases.

The structure of this study is inspired by \cite{ali2019fused}; Uses pre-trained variants of ResNet50,
InceptionV3, and Inception-ResNetV2 to obtain prediction accuracies of 98\%, 97\%, and 87\%, 
respectively \cite{ali2019fused}. This showed that ResNet50 is an effective pre-trained model for CXR COVID-19
detection. However, our study improved upon this previous work by expanding the dataset from
50 COVID-19 images \cite{ali2019fused} to 261 COVID-19 images from the COVID-19-CT-CXR \cite{peng2020covid} dataset
and by introducing bagging ensembles to further increase accuracy.
This manuscript is organized as follows: the dataset is covered in Section 2.1; data preprocessing is described in Section 2.2; Section 2.3 summarizes the network architectures used;
deep transfer learning and ensemble learning are explained in Section 2.4 and 2.5, respectively;
Section 2.6 and 2.7 outline the experimental setup and performance metrics, respectively.
Section 3 contains the Discussion and obtained results for each neural network. Section 4
summarizes the work with a Conclusion.

\section{Materials and Methods}
\label{sec:headings}

\subsection{Dataset}
In this study, 261 COVID-19 chest X-ray images have been obtained from the open-source
COVID-19-CT-CXR dataset \cite{peng2020covid}. These images were extracted from figures, associated captions,
and subfigures in COVID-19 articles from the PubMed Central Open Access (PMC-OA) subset.
Most figures found in the PMC-OA articles are compound figures, each consisting of several
subfigures often being of multiple categories such as CT, CXR, or an assortment of 26 other
scientific figures. A deep-learning model was designed to distinguish compound figures from
other figure types \cite{peng2020covid}. The designed deep-learning model is a convolutional neural network pretrained on the ImageCLEF Medical dataset  \cite{tsutsui2017data}. 

Many of the figures in the PMC-OA articles are not CT or CXR images. Therefore, a
scientific figure classifier based on the DenseNet121 architecture was trained on 500 CXR
images and 500 CT images taken from the NIH Chest X-ray and DeepLesion dataset,
respectively. DenseNet121 was pretrained on the ImageNet dataset and fine-tuned with the last
classification layer replaced with a fully connected layer using a softmax operation \cite{peng2020covid}. 
For text extraction, PMC-OA articles were downloaded through a RESTful web service.
The articles were parsed for figure numbers and captions, and figure numbers and regular
expressions were used to find where figures were cross-referenced \cite{peng2020covid}.

\begin{figure}[htp]
  \centering
  \begin{minipage}[b]{0.4\textwidth}
    \includegraphics[width=\textwidth]{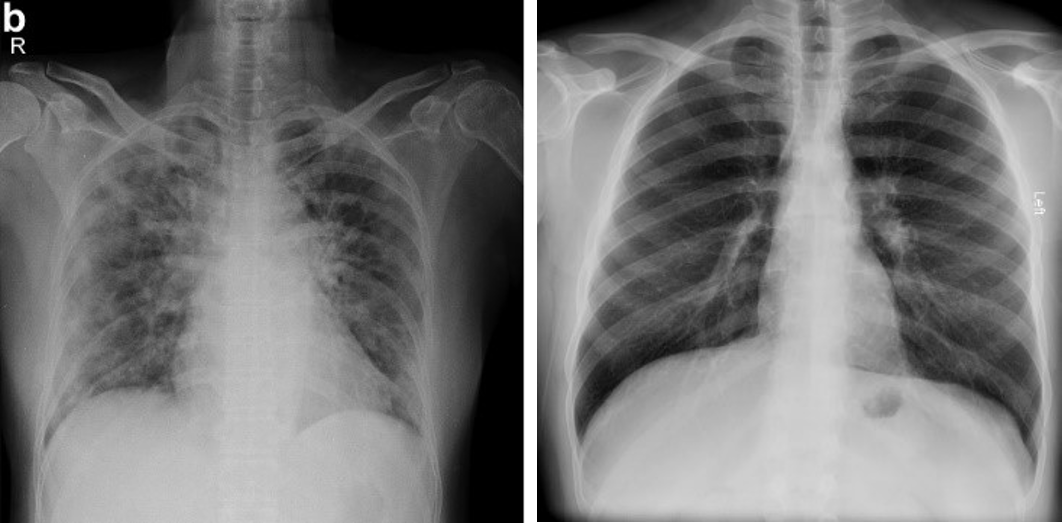}
    \caption{CXR Images of COVID-19 Cases }
  \end{minipage}
  \hfill
  \begin{minipage}[b]{0.4\textwidth}
    \includegraphics[width=\textwidth]{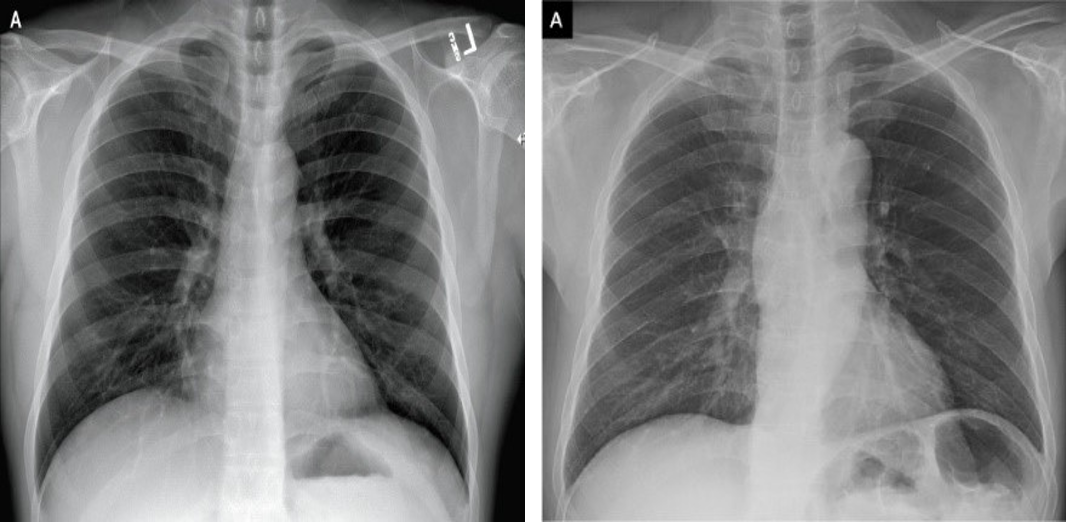}
    \caption{CXR Images of Normal Cases}
  \end{minipage}
  \label{fig:fig1-2}
\end{figure}

\subsection{Data Pre-processing}
Images from the COVID-19-CT-CXR \cite{peng2020covid} dataset are taken at different centers, with different protocols, and
with the potential for human error. To account for these differences and to artificially expand our
limited dataset, we use random x-axis and y-axis reflections each with a 50\% probability,
random rotation from 10 to -10 degrees, and a random x-axis and y-axis shear from -0.3 to 0.3
degrees.

Images from the COVID-19-CT-CXR dataset range in height from 224 to 2,703 pixels
with an average of 387.5 pixels. Image width ranges from 224 to 1,961 pixels with an average of
472.4 pixels \cite{peng2020covid}. Each image is resized to fit the input of ResNet18, ResNet50, and DenseNet201
with an image input size of 224 x 224 pixels and the input of Xception with an image input size
of 299 x 299 pixels.

\subsection{Network Models}
Three models – ResNet18 \cite{zagoruyko2016wide}, ResNet50  \cite{zagoruyko2016wide}, and DenseNet201  \cite{huang2017densely} – are pretrained on the
ImageNet dataset then fine-tuned on the COVID-19-CT-CXR dataset with the final-layers
replaced. Once fine-tuned, experiments are done to determine the robustness of individual
ensemble network combinations.

\subsubsection{ResNet}
It has been shown that deep residual networks have increasing performance as layers are scaled
up. However, these networks have a problem of diminished feature reuse known as the
vanishing/exploding gradient problem \cite{zagoruyko2016wide,bengio1994learning,hinton2012improving}. To mitigate this problem, a new architecture
called ResNet blocks was introduced to decrease depth and increase width of residual networks.
ResNets use identity connections and stochastic depth to move signals from one layer to the next
and to randomly drop layers to promote gradient flow, respectively. It is demonstrated that even
shallow wide residual networks outperform other previous deep learning networks \cite{zagoruyko2016wide}.

\subsubsection{DenseNet}
Research has shown that Dense Convolutional Networks (DenseNets) connecting each layer to
every other layer in a densely connected block alleviate the vanishing/exploding gradient
problem \cite{huang2017densely,bengio1994learning,hinton2012improving,ragb2016histogram}, reduce parameter count, and encourage feature reuse. This means that for
traditional convolutional networks with L connections between layers, a DenseNet will have
$ \frac{L(L + 1)}{2} $ connections. Both ResNet and DenseNet share a similar characteristic in that they
shorten the distance from earlier layers to later layers \cite{huang2017densely}.

\begin{table}[htp]
\centering
\captionsetup{justification=centering}

\begin{tabular}{|c|c|c|c|}
Network     & Depth & Parameters (Millions) & Image Input Size  \\ \hline
ResNet18    & 18    & 11.7                  & 224 x 224         \\
ResNet50    & 50    & 25.6                  & 224 x 224         \\
DenseNet201 & 201   & 20                    & 224 x 224        
\end{tabular}

\caption{Network Models}
\end{table}

\subsection{Deep Transfer Learning}
Data dependence, where deep learning has a huge dependence on large-scale datasets, is a serious problem. With an improvement in deep learning’s potential as a function of layer count, a linear relationship also exists between model scale and dataset size. Insufficient training data is eminent in bioinformatics datasets where data collection is an expensive and complicated process \cite{tan2018survey}.
In the domain of bioinformatics, large-scale and well-annotated datasets are costly to produce, thus limiting the size of the dataset. Transfer learning allows for the use of datasets that are not identically distributed to the test dataset. Expounding further, an unrelated dataset can be used to train the early layers of a deep neural network responsible for the identification of edges, blobs, and colors. Then, the intended dataset – in this instance, COVID-19-CT-CXR – is used to train the neural network to identify complex features utilizing those earlier edges, blobs, and colors \cite{tan2018survey}.
 A Survey on Deep Transfer Learning, Chuanqi Tan and colleagues characterize deep transfer learning into four distinct categories: instance-based, mapping-based, network-based, and adversarial-based \cite{tan2018survey}. For the purposes of this paper, network-based deep transfer learning is utilized. This involves the reuse of a partial-network (front-layers) pre-trained on a source domain then transferred to be retrained on the target domain. The front-layers are treated as a versatile feature extractor that is useful for other learning tasks. 
Studies the transition from generalized to specific features from the early-layers to the final-layers \cite{yosinski2014transferable}. It is stated that first-layer features tend to learned features imitate the Gabor filters and color blobs \cite{yosinski2014transferable}. is an in-depth study between network structure and transferability. It concludes that ResNet models are good candidates for deep transfer learning \cite{tan2018survey,yosinski2014transferable}.
The transfer learning models used in this paper were pre-trained on the ImageNet dataset used in the ImageNet Large-Scale Visual Recognition Challenge (ILSVRC) consisting of over a million images and a thousand class labels. For each network, the last three layers – fully connected, softmax, and classification layer – are replaced with the weights randomly initialized. The fully connected, softmax, and classification layers are responsible for identifying those features specific to COVID-19 detection.

\begin{figure}[htbp]
\centerline{\includegraphics[width=0.6\textwidth]{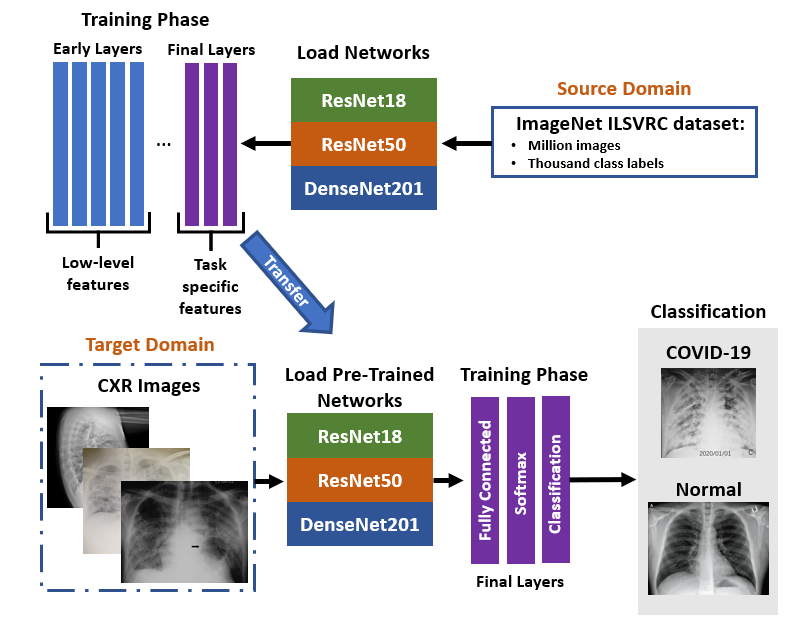}}
\caption{Network-Based Deep Transfer Learning for the Prediction of COVID-19 Patients Using CXR Images}
\label{fig:fig3}
\end{figure}

\subsection{Ensemble Learning}
\begin{figure}[htbp]
\centerline{\includegraphics[width=0.6\textwidth]{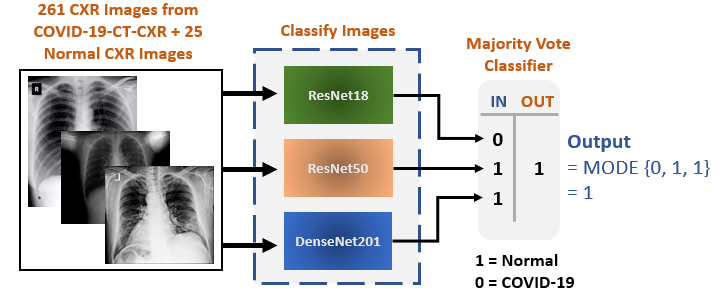}}
\caption{Bagging Approach \cite{opitz1999popular} to Ensemble Learning using a Majority Vote Classifier}
\label{fig:fig4}
\end{figure}

\begin{figure}[!htp]
\centerline{\includegraphics[width=0.5\textwidth]{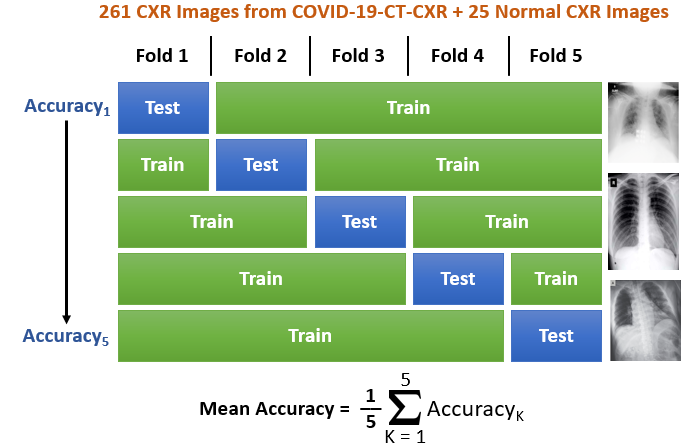}}
\caption{Visual Display of K-fold Cross-validation (k = 1-5)}
\label{fig:fig5}
\end{figure}

An ensemble is a set of classifiers whose predictions are combined. Research has shown that an ensemble tends to be more accurate than any individual classifier for various imaging applications \cite{ali2019deep,ali2019fused,narayanan2019performance}. Two popular methods of ensemble learning are bagging classifiers and boosting classifiers. Bagging classifiers train each classifier on varying subsets of the target dataset. This is particularly effective in “unstable” neural networks where small changes in the training set results in noticeable changes in test set predictions \cite{opitz1999popular}. Moreover, each network in the ensemble may have an innate propensity to predict different subsets of the dataset more accurately, further supporting the bagging classifier’s effectiveness. As a result, bagging classifiers generally always outperform any individual classifier in the ensemble \cite{opitz1999popular}.
Boosting classifiers are a group of methods which involve altering the input dataset to the training phases of proceeding networks in an ensemble based on the performance of preceding networks. This is the practice of tailoring the input datasets of networks such that an optimal dataset representation is achieved that is non-reliant on input variance across the ensemble classifier. This ensemble method can vastly outperform bagging classifiers, but it has the potential to perform worse than an individual network in the classifier. Boosting classifier performance is heavily dependent on features of the target dataset \cite{opitz1999popular}.
For this experiment, the bagging approach is adopted to maximize performance. Each network is trained using k-fold cross-validation. The mean accuracy is measured across each fold of the trained models as seen in Figure \ref{fig:fig5}. The output classification of the three selected networks are then grouped into a majority vote classifier ensemble neural network. A majority vote classifier takes the mode of the binary classifications as seen in Figure \ref{fig:fig4}.

\subsection{Experimental Setup}
The transfer learning models were trained using the MATLAB programming language. CNN models (ResNet18, ResNet50, and DenseNet201) were pretrained on a subset of the ImageNet dataset which is used in the ImageNet Large-Scale Visual Recognition Challenge (ILSVRC) consisting of over a million images and a thousand class labels. For each network, the last three layers – fully connected, softmax, and classification layer – are replaced with the weights randomly initialized. Each network was trained for 15 epochs with a batch size and learning rate of 8 and 0.00005, respectively. As a cross-validation method, k-fold is chosen as a strategy to combat the limited data samples. The dataset split was 20\% test set and 80\% training set in each fold. K-fold cross-validation is used to ensure that each observation from the raw dataset can appear in both the training and testing set. Results were obtained using 5 different k values (1-5). Then each network is combined into a bagging ensemble neural network. This bagging ensemble uses a majority vote classifier which takes the mode of each output of the combined neural networks as seen in Figure \ref{fig:fig4}. It is expected that a bagging ensemble will outperform any individual network model.

\subsection{Performance Metrics}
\begin{figure}[htbp]
\centerline{\includegraphics[width=0.4\textwidth]{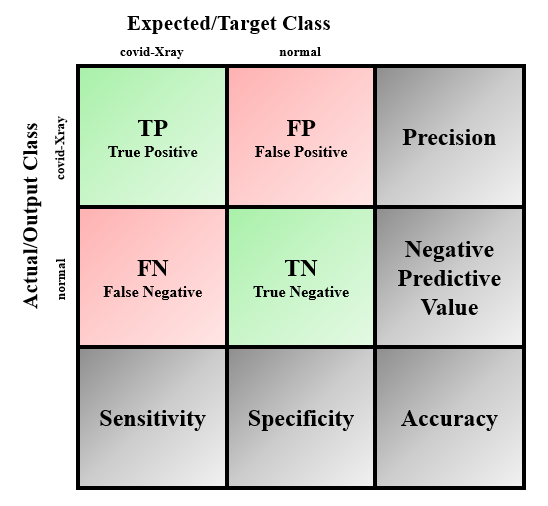}}
\caption{Confusion Matrix for Applied Deep Learning}
\label{fig:fig6}
\end{figure}

TP, FP, TN, and FN given in Figure \ref{fig:fig6} represent True Positive, False Positive, True Negative, and False Negative percentages, respectively. Diagnostic testing necessitates a low False Negative (FN) rate or high sensitivity often at the risk of increasing the False Positive (FP) rate. These values are used in Equations (1) - (7) as evaluation metrics for the effectiveness of each neural network:

\begin{enumerate}
	\item $ Accuracy = \frac{(TN + FP)}{(TN + TP + FN + FP)} $
	\item $ Sensitivity = \frac{TP}{(TP + FN)} $
	\item $ Specificity = \frac{TN}{(TN + FP)} $
	\item $ Precision = \frac{TP}{(TP + FP)} $
	\item $ Negative Predictive Value = \frac{TN}{(TN + FN)} $
	\item $ F1 Score = 2\left[\frac{Precision * Recall}{Precision + Recall}\right] $
	\item $ False Positive Rate = \frac{FP}{(FP + TN)} $
\end{enumerate}

Free-response receiver operating characteristic (FROC) curves allows for the evaluation of neural network performance at all classification thresholds.  A classification threshold is the neural network’s resistance against labeling an image as positive (COVID-19); further expounding, the decision threshold is a neural network’s likelihood of classifying an image as negative (Normal). For medical diagnostic purposes, it is imperative that all positive cases are correctly classified. Therefore, a lower decision threshold is preferred. At each decision threshold, the sensitivity or true positive rate (TPR) as found in Equation (2) is plotted in the y-axis against the ‘average FPs per image’ or false positive rate (FPR) as found in Equation (7). As the threshold is lowered, both the TPR and FPR increase or remain constant. 
Area under the ROC curve (AUC) aggregates the performance of the neural network on all decision thresholds. This represents an average performance for the neural network. The closer the AUC is to 1.0, the better the network is at distinguishing between positive and negative classes.

\section{Results and Discussion}
In this study, ResNet18, ResNet50, and DenseNet201 have been trained on chest X-ray images from the COVID-19-CT-CXR dataset. Each network has a confusion matrix, FROC curve, and their 95\% confidence intervals demonstrating their performance on several metrics. ResNet18 achieved the highest precision (99.6\%) of each of the trained networks while ResNet50 achieved the highest sensitivity (99.6\%) of each of the trained networks. The two networks were combined in a bagging ensemble alongside DenseNet201 to achieve a network with both a high precision and sensitivity of 99.6\% and 100\%, respectively. Our evaluation also shows that the bagging ensemble method achieves the best performance for classification (overall accuracy: 99.7\%, 95\% confidence interval: 87.75\%–94.22\%). The lowest performance values were yielded by DenseNet201 with a sensitivity and negative predictive value of 98.5\% and 85.2\%, respectively. Across the three neural networks, only one CXR images was a false positive between two of them, and no CXR images were false negatives implying that ResNet18 and DenseNet201 tend to identify disparate features in labeling false negatives – a desirable characteristic.

\subsection{Figures and Tables}

\begin{figure}[!tbp]
  \centering
  \begin{minipage}[b]{0.4\textwidth}
    \includegraphics[width=\textwidth]{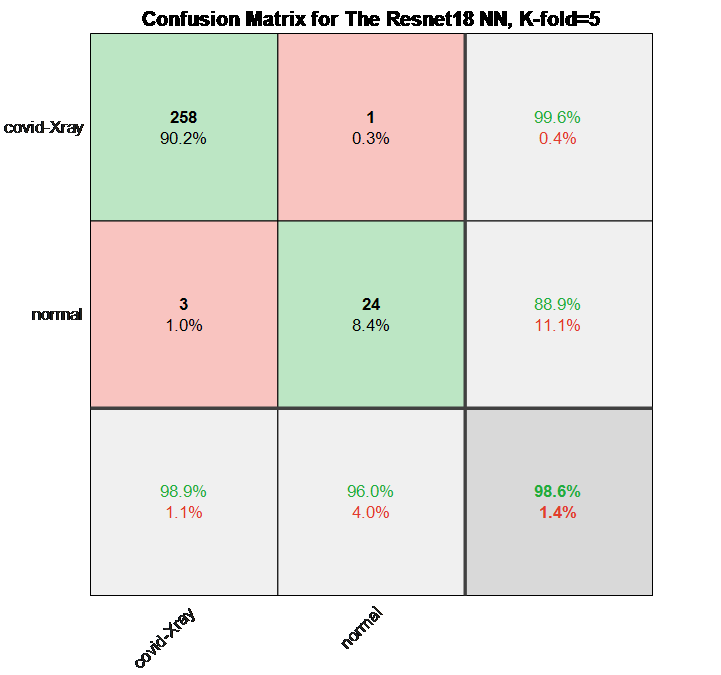}
    \caption{Confusion Matrix for ResNet18}
  \end{minipage}
  \hfill
  \begin{minipage}[b]{0.4\textwidth}
    \includegraphics[width=\textwidth]{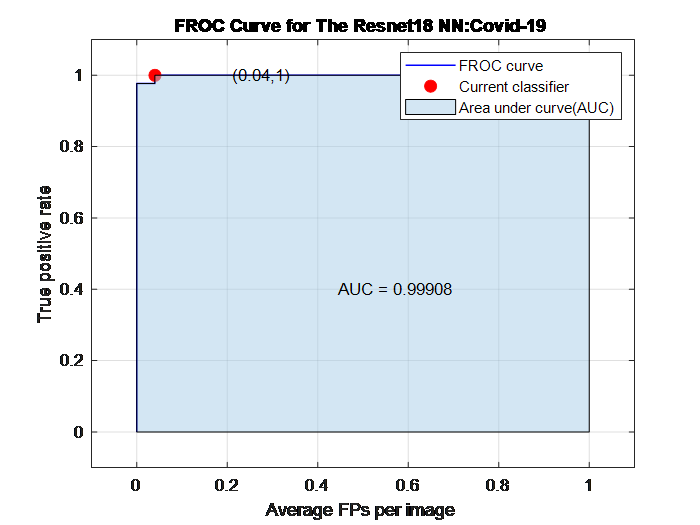}
    \caption{FROC Curve for ResNet18}
  \end{minipage}
\end{figure}

\begin{figure}[!tbp]
  \centering
  \begin{minipage}[b]{0.4\textwidth}
    \includegraphics[width=\textwidth]{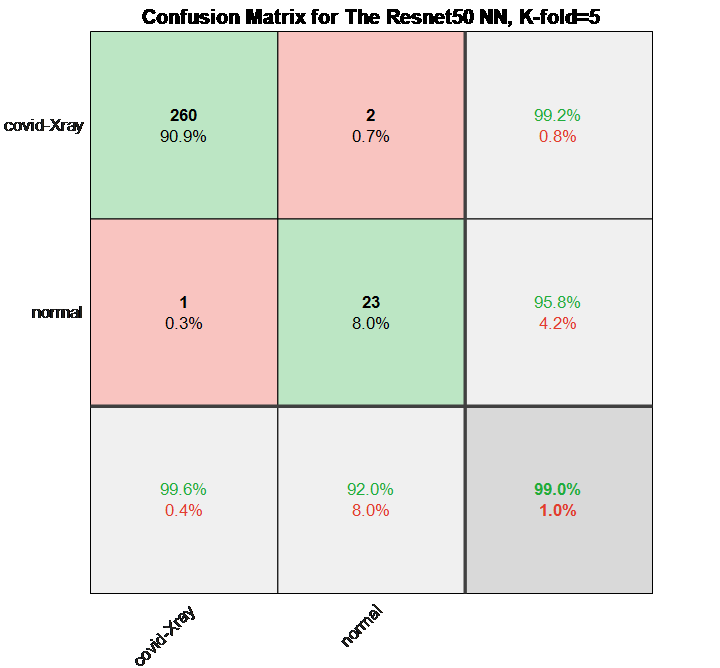}
    \caption{Confusion Matrix for ResNet50}
  \end{minipage}
  \hfill
  \begin{minipage}[b]{0.4\textwidth}
    \includegraphics[width=\textwidth]{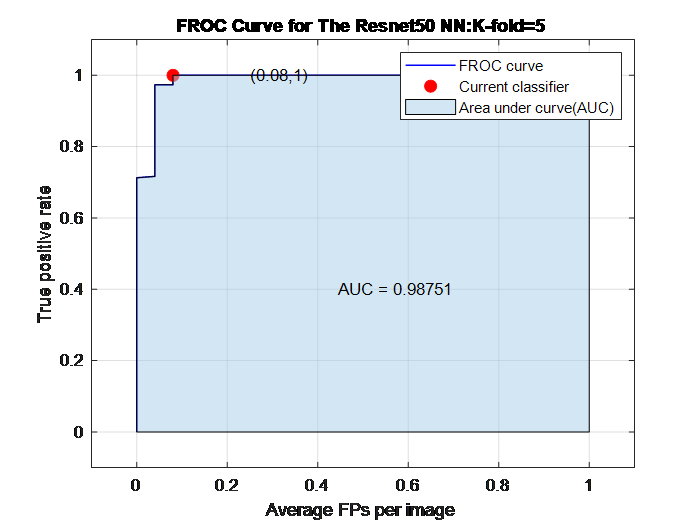}
    \caption{FROC Curve for ResNet50}
  \end{minipage}
\end{figure}

\begin{figure}[!tbp]
  \centering
  \begin{minipage}[b]{0.4\textwidth}
    \includegraphics[width=\textwidth]{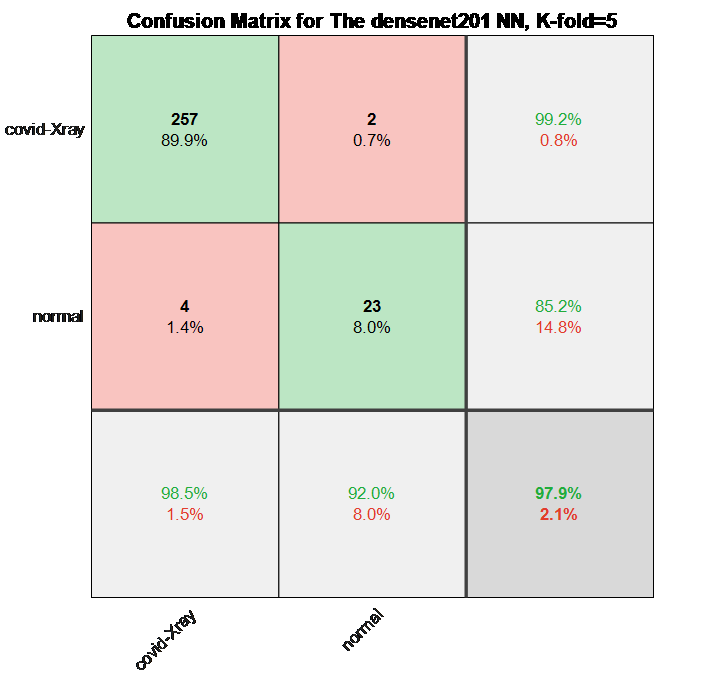}
    \caption{Confusion Matrix for DenseNet201}
  \end{minipage}
  \hfill
  \begin{minipage}[b]{0.4\textwidth}
    \includegraphics[width=\textwidth]{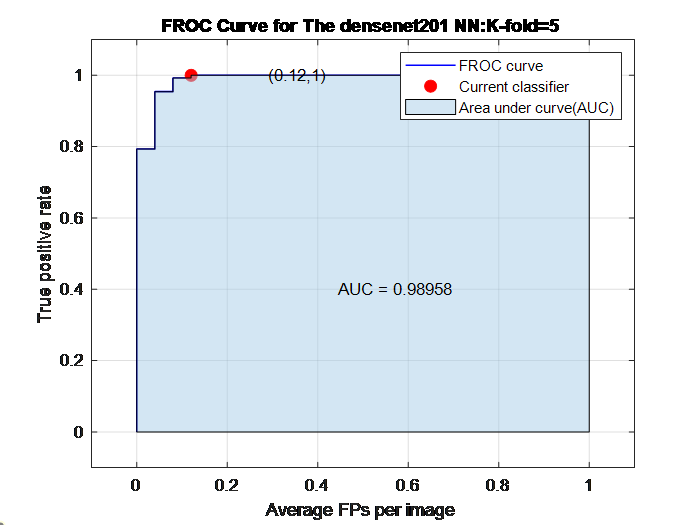}
    \caption{FROC Curve for DenseNet201}
  \end{minipage}
\end{figure}

\begin{figure}[!tbp]
  \centering
  \begin{minipage}[b]{0.4\textwidth}
    \includegraphics[width=\textwidth]{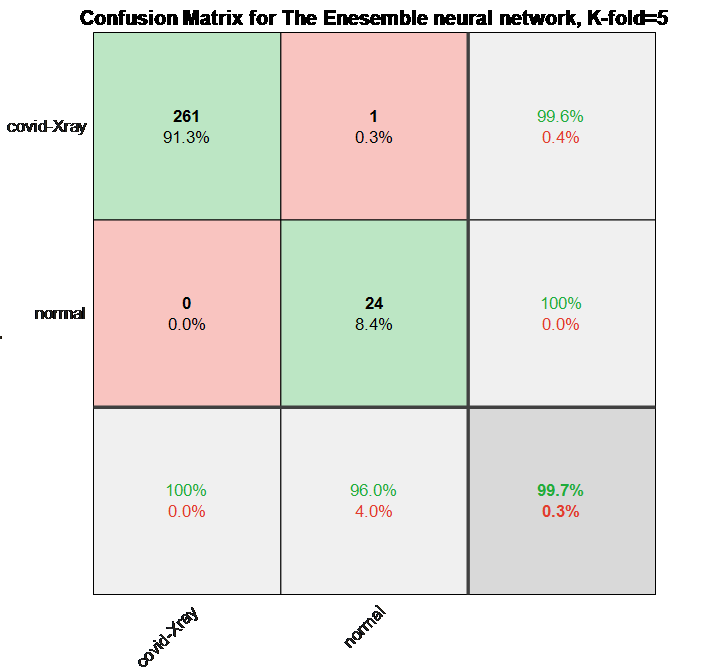}
    \caption{Confusion Matrix for Ensemble Neural Network}
  \end{minipage}
  \hfill
  \begin{minipage}[b]{0.4\textwidth}
    \includegraphics[width=\textwidth]{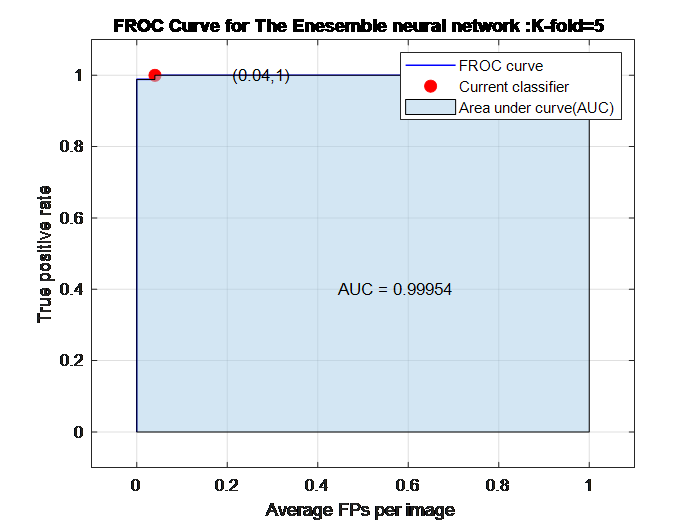}
    \caption{FROC Curve for Ensemble Neural Network}
  \end{minipage}
\end{figure}

\begin{table}[htp]
\centering

\begin{tabular}{|c|c|c|c|c|c|c|c|}
Neural Network & TP & FP & TN & FN & Accuracy & AUC & Confidence Interval (95\%)  \\ \hline
ResNet18 & 90.2\% & 0.3\% & 8.4\% & 1\% & 98.6\% & 0.99908 & 84.69 – 91.82 \\
ResNet50 & 90.9\% & 0.7\% & 8.0\% & 0.3\% & 99.0\% & 0.98751& 86.53 – 93.12 \\
DenseNet201 & 89.9\% & 0.7\% & 8.0\% & 1.5\% & 97.9\% & 0.98958 & 86.98 – 93.69\\
Ensemble NN & 91.3\% & 0.3\% & 8.4\% & 0\% & 99.7\% & 0.99954 & 87.75 – 94.22 \\
\end{tabular}

\caption{Confusion Matrix Values of Each Neural Network (k = 1-5)}
\end{table}

\section{Conclusion}
Early, relatively inexpensive, and sanitary detection of COVID-19 is of upmost importance during the current stages of the Coronavirus pandemic. By training deep transfer learning models and combining their outputs into a majority vote ensemble, we propose an effective binary classifier for CXR imagery as a first-line triage tool for hospitals and radiology services worldwide. Through ResNet50’s exceptional performance alongside ResNet18 and DenseNet201’s complementary image detection characteristics, an ensemble neural network with only one mislabeled CXR image of amongst a total of 286 CXR images is achieved. This study gives insight into how ensemble neural network architectures can be used to increase performance, and it is believed that this work can contribute to lessening the burden on radiology services. Further research can be performed by experimenting with new convolutional neural network architectures or with different ensemble techniques such as the boosting approach; additionally, there is potential to extend this work beyond a binary classifier into nuanced multi-outputs (viral pneumonia,  inflammatory diseases, etc.).

\bibliographystyle{unsrt}  
\bibliography{references}  



\end{document}